\documentclass[aps,pra,twocolumn,groupedaddress, showpacs, showkeys]{revtex4}
\usepackage{graphicx}
\usepackage{graphics}
\usepackage{color}      
\usepackage{subfigure}

\newcommand{\bfrt}{({\bf{r}},t)}

\begin{document}


\definecolor{AsGreen}{rgb}{0.3,0.8,0.3} 
\definecolor{AsRed}{rgb}{0.6,0.0,0.0}
\definecolor{AsLightBlue}{rgb}{0,0.6,0.6}
\definecolor{AsPurple}{rgb}{0.6,0,0.6}
\definecolor{mypurple}{rgb}{0.5, 0.14, 0.6}

\hyphenpenalty=5000
\tolerance=1000

\title{Isotropic vortex tangles in trapped atomic Bose-Einstein condensates via laser
  stirring}
\author{A.~J. Allen}
\author{N.~G. Parker}
\author{N.~P. Proukakis}
\author{C.~F. Barenghi}
\email{carlo.barenghi@ncl.ac.uk}
\affiliation{Joint Quantum Centre (JQC) Durham-Newcastle, School of Mathematics and Statistics, 
Newcastle University, Newcastle upon Tyne, NE1 7RU, England, UK.}
\date{\today}
  \begin{abstract}
    The generation of isotropic vortex configurations in trapped atomic Bose-Einstein condensates
    offers a platform to elucidate quantum turbulence on mesoscopic scales.
We demonstrate that a laser-induced obstacle moving in a figure-eight path within the condensate provides a
simple and effective means to generate an isotropic three-dimensional vortex tangle due to its minimal net transfer of angular momentum to the condensate.  Our characterisation of vortex structures and their isotropy is based
on projected vortex lengths and velocity statistics obtained numerically via 
the Gross-Pitaevskii equation.  Our methodology provides a possible experimental
route for
generating and characterising vortex tangles and quantum turbulence in atomic Bose-Einstein condensates. 
  \end{abstract}
  \maketitle
  Vortices in ordinary (classical) fluids, as well as quantum fluids, characterise turbulent
  flow~\citep{frisch_95,barenghi_donnelly_01}.  
  Turbulence in classical fluids has been intensely studied in many branches of physics and 
  engineering over a prolonged period.  Characterising turbulence and understanding its dynamics is 
  one of the key goals of these fields. Homogeneous, isotropic turbulence is the 
  benchmark to understand vortex dynamics away from boundaries.  Quantum fluids, such as superfluid He and atomic
  Bose-Einstein condensates (BECs), where the circulation is quantized and viscosity is
  absent, open up the possibility
  of a context in which to study turbulence which is simpler than in ordinary fluids.  Large vortex tangles have been created experimentally
  in superfluid He for this purpose.  The investigation of the properties of such systems has
  revealed, for certain parameter regimes,
  the emergence of classical-like behaviour (such as the Kolmogorov
  scaling~\citep{maurer_tabeling_98,salort_baudet_10,barenghi_lvov_13} of the energy spectrum for homogeneous isotropic turbulence)
 from the dynamics of elementary quantum vortices.  

\begin{figure}[h!]
    \centering{
      \includegraphics[clip,width = 4.1cm]{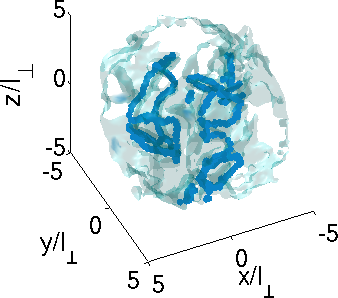}   
       \includegraphics[clip,width = 4.3cm]{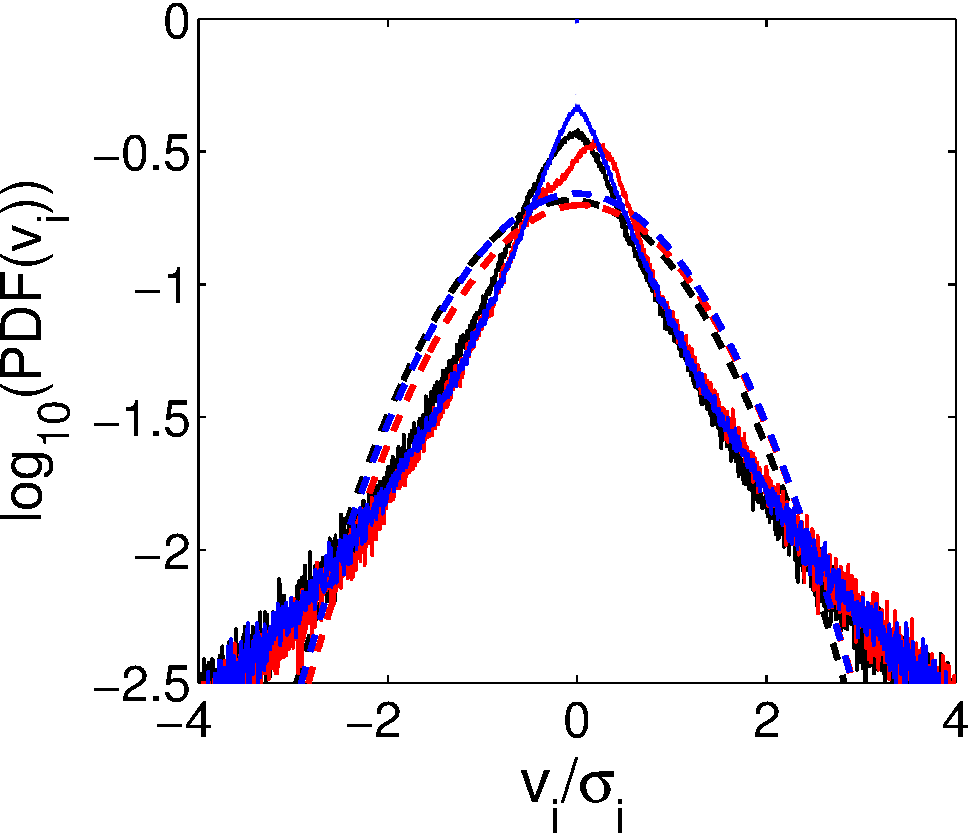}  
    }
    \caption{Tangle of few interacting vortices (left) exhibiting isotropic non-Gaussian
      flow statistics (right).  Left: Density isosurface generated by stirring a spherically
      symmetric condensate along one plane in a figure-eight path (at time $t\approx
      t_{\rm{stir}}=17.1\omega^{-1}$ when the stirrer has just been removed after
      approximately two oscillations), with the 
    vortex cores visualised by the dark blue dots.  Right: PDFs of
    velocity components $v_x$ (black), $v_y$ (red) and  $v_z$ (blue) at 
    time $t\approx t_{\rm{stir}}$.  The non-Gaussian nature of the
    turbulent velocity field is made apparent by the large deviation of the PDFs from the
    corresponding gPDFs, calculated by
    Eq.~(\ref{gpdfs}) (dashed lines).
}
\label{fig:allen_fig1}
\end{figure}

Usually terms like ``turbulence'' and ``vortex tangles'' refer to disordered fluid systems
containing vortices and eddies in which a huge range of lengthscales and timescales are
excited; scaling laws therefore can be identified.  Unlike ordinary fluids and superfluid
helium, atomic BECs are relatively small, in the sense that there is not a large
separation of lengthscales between the vortex core size, the average intervortex
separation and the system size~\citep{allen_parker_13}.  An important question which
should be addressed in this context, is whether a relatively small vortex configuration
exhibits turbulent properties, or it is simply chaotic.  A first step in addressing this
question is to demonstrate a technique for generating a few interacting vortices (see
Fig.~\ref{fig:allen_fig1} (left)) that give
isotropic flow statistics (see Fig.~\ref{fig:allen_fig1} (right)), which is the main result of this paper.

  The experimental realisation of BECs
  ~\citep{anderson_ensher_95,davis_mewes_95,bradley_sackett_97} has
  opened up the possibility of experimentally
  generating a turbulent tangle of vortices in a highly controllable system.  
  Vortex lattices~\citep{madison_chevy_00,hodby_hechenblaikner_01,abo-shaeer_raman_02,aboshaeer_raman_01,madison_chevy_01,raman_aboshaeer_01} and collections of vortex dipoles~\citep{neely_samson_10}, as well as multiply-charged~\citep{shin_saba_04} and multi-component~\citep{matthews_anderson_99,anderson_haljan_00,schweikhard_coddington_04} 
 vortices can be created, and their two dimensional (2D) column distributions can even be imaged in real-time~\citep{freilich_bianchi_10}.
 With regards to quantum turbulence, a landmark experiment recently generated a
 small tangle of vortices in a trapped weakly-interacting BEC 
  through the combination of rotation and an external oscillating
  perturbation~\citep{henn_seman_09, seman_henn_11, shiozaki_telles_11}. Several theoretical works have studied the
  statistical properties of vortex tangles in three-dimensional (3D) superfluid
  systems~\citep{nore_abid_07,berloff_svistunov_02,yepez_vahala_09,white_barenghi_10}.
  Another proposed method to generate a vortex tangle combines rotations about two
  different axes~\citep{kobayashi_tsubota_07}.  

Atomic BECs are typically inhomogeneous
  because of the harmonic potentials which are used to confine them (although there are
  ongoing efforts to minimize the effects of inhomogeneity via the creation of `box-like'
  condensates~\citep{gaunt_schmidutz_13}); because of the small number of vortices generated,
  the prevailing vortex configuration is unlikely to be isotropic if it contains only a few
  vortices.  A summary of the features and
  main unresolved issues in such systems can be found in~\citep{allen_parker_13}.  

   It is well known that an obstacle moving through a superfluid will nucleate
  vortices above a critical speed~\citep{frisch_pomeau_92}.  In a BEC
  such an obstacle can be generated via an incident blue-detuned laser beam, which
  induces a localised repulsive potential through the optical dipole force.  Deflection
  of the laser beam can then be performed to move the obstacle over time.  This technique, termed 
  ``laser stirring''~\citep{caradoc-davis_ballagh_99,raman_kohl_99,raman_aboshaeer_01,kevrekidis_frantzeskakis_08}, has proven to be an efficient way
  of generating large numbers of 
  vortices both experimentally ~\citep{raman_aboshaeer_01,neely_samson_10,raman_kohl_99,kevrekidis_frantzeskakis_08} and theoretically~\citep{caradoc-davis_ballagh_99,white_barenghi_12,reeves_anderson_12,reeves_billam_13}.  In particular, laser stirring in a circular path has led to the formation of vortex 
  lattice
  states~\citep{raman_aboshaeer_01}
  due to its imparting of angular momentum to the condensate (the ground state of a
  superfluid with sufficient angular momentum being a vortex
  lattice~\citep{donnelly_91,pethick_smith_book_02,kevrekidis_frantzeskakis_08}).

  In this paper we show that laser stirring in a figure-eight pattern in a two
  dimensional plane can be used to generate a fully three dimensional isotropic vortex
  tangle in a trapped BEC, a vortex configuration suitable to study quantum turbulence.
  While this stirring is an efficient means of generating vortices it imparts minimal angular
  momentum to the condensate, thus favouring the generation of a tangle of vortices rather than a vortex lattice.
  Following
  cessation of the stirring, the tangle decays isotropically.  Since velocity
  statistics~\citep{paoletti_fisher_08} and vortex length~\citep{donnelly_91} are
  routinely used to measure tangle dynamics in He experiments of quantum turbulence, we
  apply these measures to characterise the vortex dynamics.

\begin{figure}[h!]
    \begin{minipage}{2.4cm}
      \includegraphics[width = 2.3cm, height = 2.3cm]{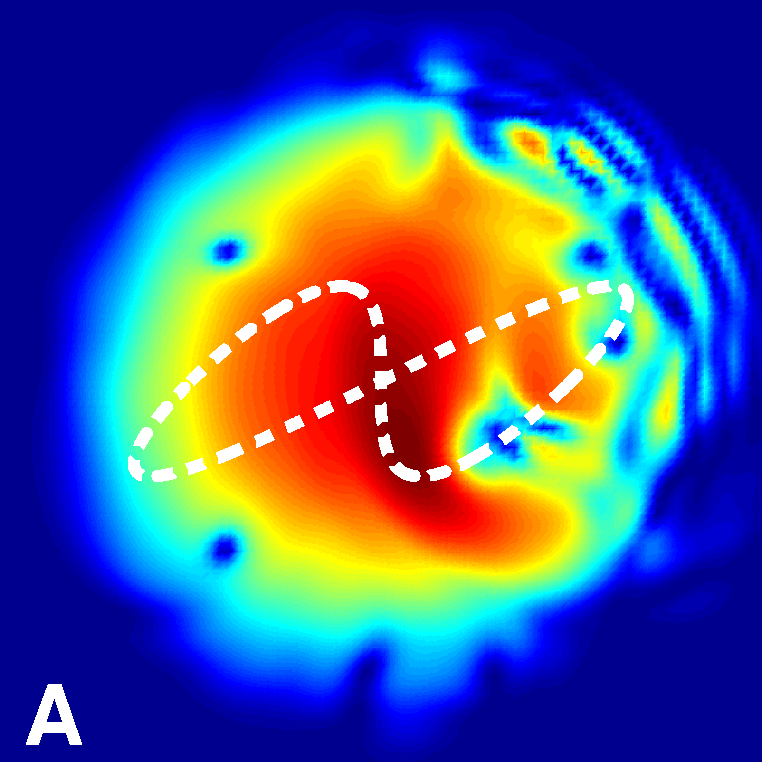}
    \end{minipage}
    \begin{minipage}{2.4cm}
      \includegraphics[width = 2.3cm, height = 2.3cm]{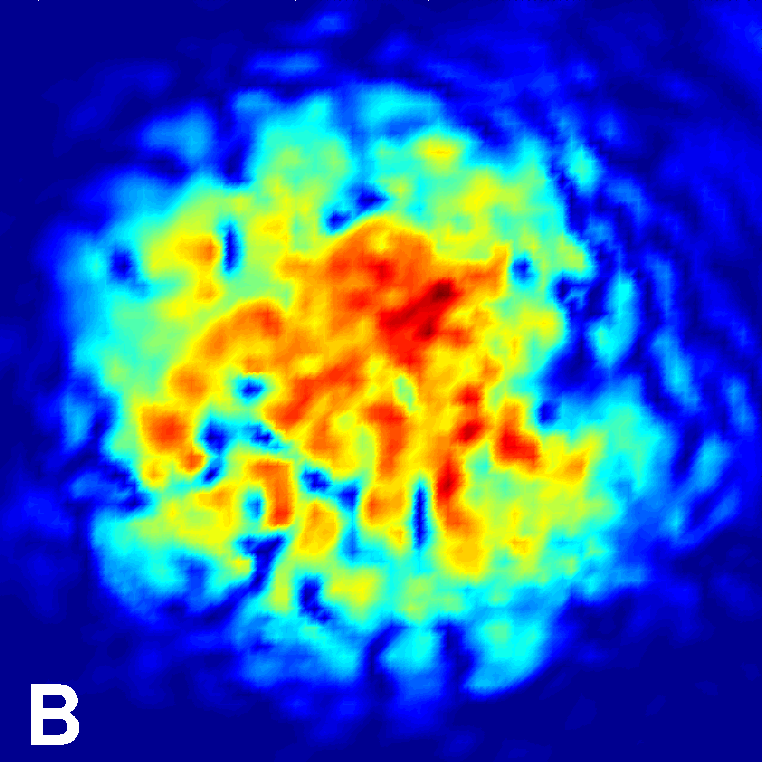}
    \end{minipage}
    \begin{minipage}{2.4cm}
      \includegraphics[width = 2.3cm, height = 2.3cm]{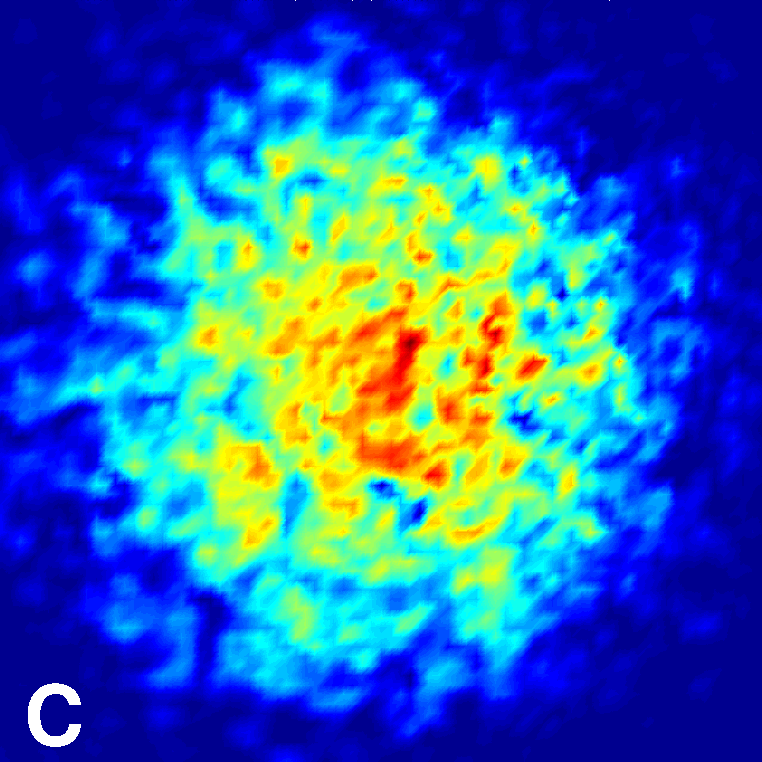}
    \end{minipage}
     \begin{minipage}{0.5cm}
       \vspace{-0.2cm}
       \includegraphics[height = 2.5cm]{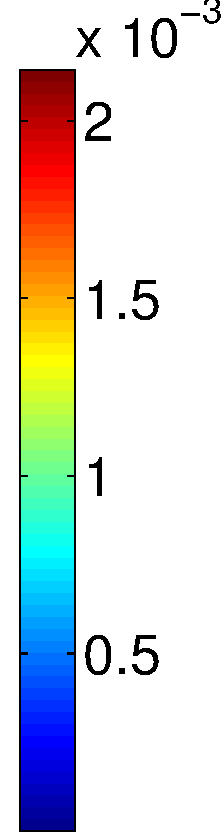}
      \end{minipage}\\
       \vspace{0.2cm}
    \begin{minipage}{2.4cm}
      \includegraphics[width = 2.3cm, height = 2.3cm]{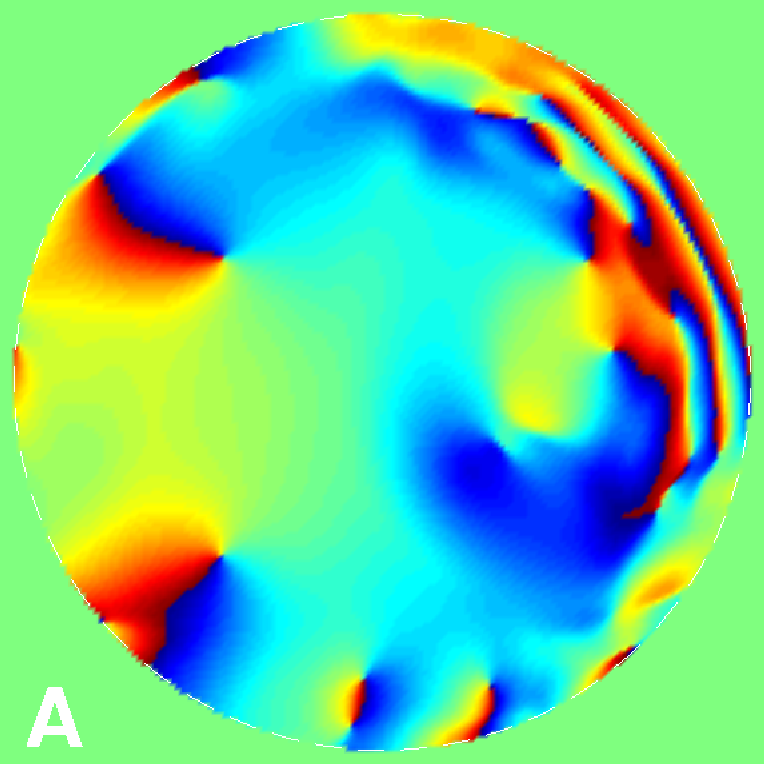}
    \end{minipage}
    \begin{minipage}{2.4cm}
      \includegraphics[width = 2.3cm, height = 2.3cm]{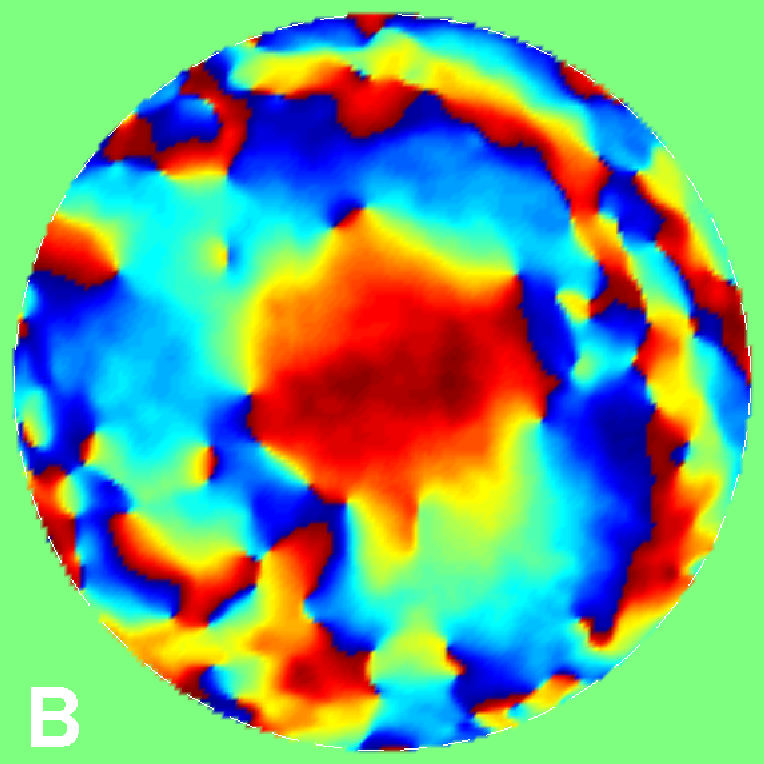}
    \end{minipage}
    \begin{minipage}{2.4cm}
      \includegraphics[width = 2.3cm, height = 2.3cm]{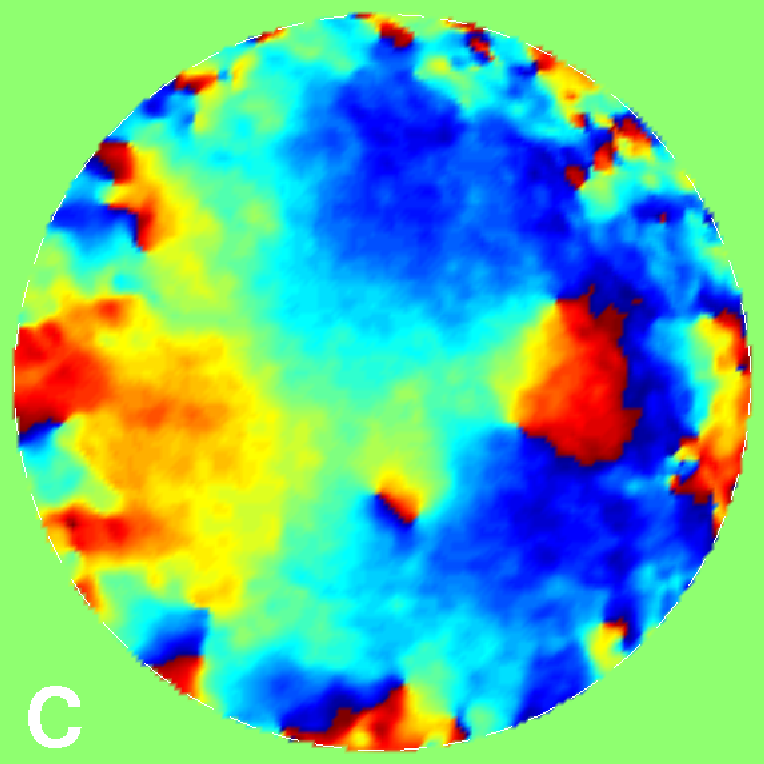}
    \end{minipage}
     \begin{minipage}{0.5cm}
       \includegraphics[height = 2.4cm]{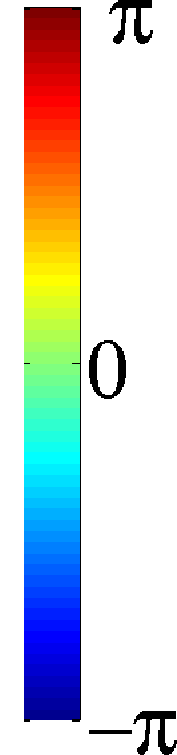}
      \end{minipage}
     \\

       \centering{
         \vspace{0.2cm}
   \includegraphics[scale = 0.295,clip]{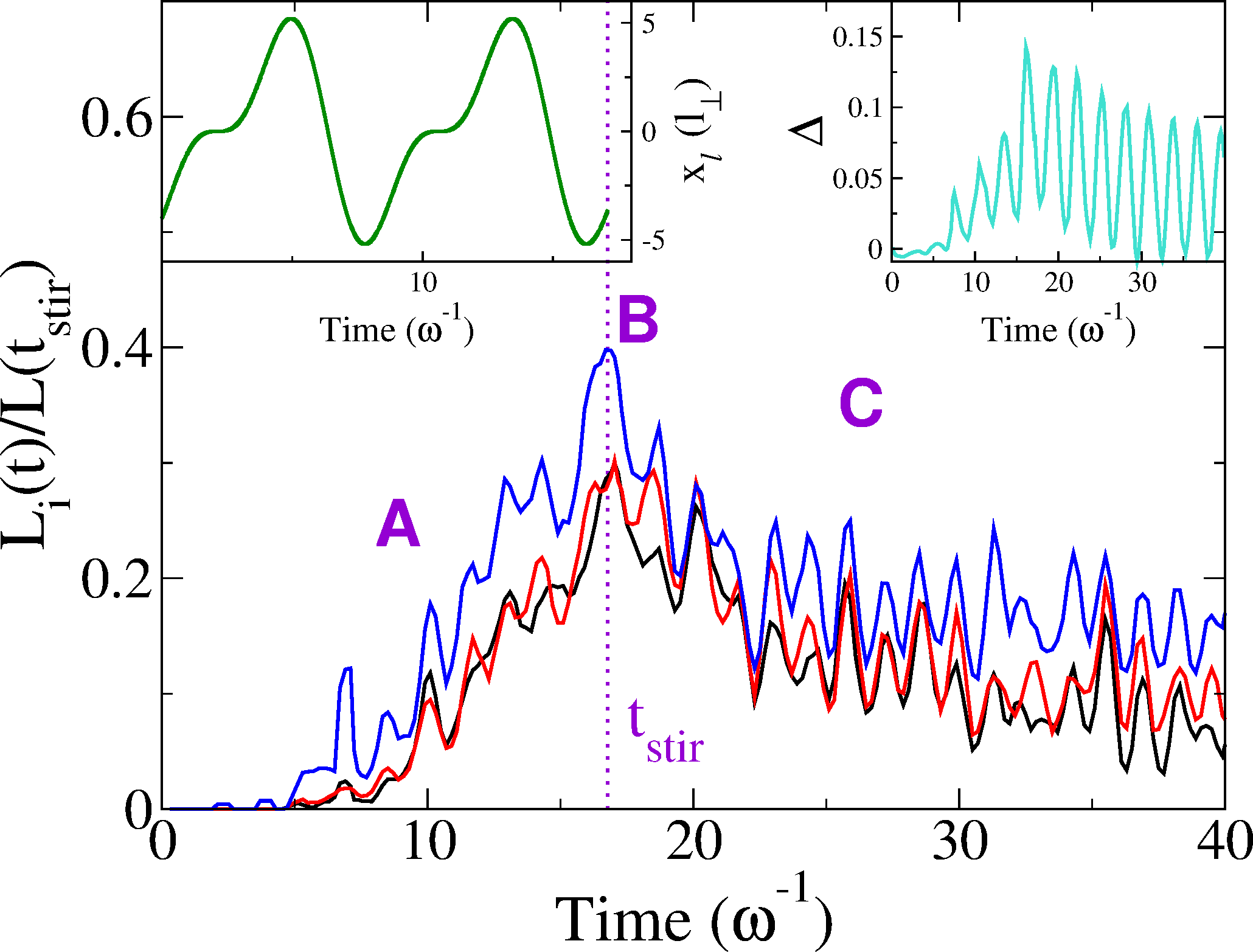}}
   \caption{(Color online) Top: Density images of the condensate at $z =0$ and $x,y = [-8,8]l_\perp$ at times
     $\omega t = 
       9.1$ (A) and $17.1$ (B) and $24.9$ (C) (also marked on the lower graph where (B)
       refers to the snapshot from Fig.~\ref{fig:allen_fig1} at time $t={t_{\rm{stir}}}$).  Regions of high (brown/red) and low
        (blue) density are apparent as indicated by the colorbar.  Density is measured in units of $l_\perp^{-3}$.  The figure-eight path of the obstacle 
    is shown by the white dashed line in (A).  Intermediate: Corresponding phase snapshots.  Singly charged vortices
    correspond to $2\pi$ phase singularities.  To avoid phase artefacts at very low density, the phase is truncated
    beyond $R_{\rm{TF}}$. Bottom:  Measured projected vortex lengths $L_i$, vs. time in the $x$ (black),
  $y$ (red) and $z$ (blue) directions.
 Inset (left):  Trajectory of laser in $x-$direction, $x_l$, up to time $t_{\rm{stir}}$.
 Inset (right):  The displaced density $\Delta$ versus time, where the displaced density is the difference in the norm between a
  condensate containing no vortices, $\phi_1$, and the stirred condensate, $\phi_2$, i.e. $ \Delta = \int_{\mathcal{V}} |\phi_1|^2 dV
-\int_{\mathcal{V}}|\phi_2|^2 dV$, where ${\mathcal{V}} = (4\pi/3)R_{\rm{cut}}^3$, with $R_{\rm{cut}} = 0.78R_{\rm{TF}}$.}
  \label{fig:allen_fig3}
    \end{figure}

  Our analysis is based on numerical simulations of the 3D Gross-Pitaevskii
  equation~\citep{pitaevskii_stringari_book_03,pethick_smith_book_02}:
    \begin{equation}
      i \hbar \frac {\partial \phi \bfrt}{\partial t} = \left( -\frac{\hbar ^2}{2m}\nabla^2
      + V \bfrt + g|\phi\bfrt|^2 - \mu \right)\phi \bfrt,
     ~\label{eqn_gpe}
    \end{equation}
    an accurate model for vortical structures in the limit of zero temperature and
    weak interactions, which describes the condensate by a 
  macroscopic wavefunction $\phi \bfrt$.  This equation is commonly expressed hydrodynamically
   using a transformation  of the form $\phi \bfrt=|\phi \bfrt| e^{i\theta\bfrt}$,
   where $\theta \bfrt$ is the phase and the superfluid velocity is identified as
   ${\mathbf{v}} \bfrt=({\hbar}/{m}) \nabla \theta \bfrt$.  
  The interatomic interactions are parametrized by $g=4\pi \hbar^2 a_s/m $, where $a_s$ is
  the $s-$wave
  scattering length and $\mu$ is the condensate chemical potential.  The external
  potential acting on the BEC is of the form $V \bfrt = m \omega^2 r^2/2  +
  V_{\mathrm{l}}(x,y,t)$.   The first term represents a spherically symmetric harmonic trap, of frequency
  $\omega$, used to confine the gas.  The second term represents a Gaussian time-dependent
  laser-induced potential, uniform along $z$:
  \begin{equation}
    V_{\mathrm{l}}(x,y,t) =  V_0{\mathrm{exp}}{\left[-\frac{(x - x_l(t))^2 +
    (y + y_l(t))^2}{d^2}\right]}.
  \end{equation}
  The amplitude $V_0$ is proportional to the intensity of the stirring laser beam~\citep{adams_riis_97,pethick_smith_book_02}, $x_l(t)$ and
  $y_l(t)$ are the positions of the center of the stirrer at time $t$, and $d$ is the beam's width.   For a figure-eight stirring path, the time dependent coordinates of the obstacle are given by 
  $\left(x_l(t),y_l(t)\right)=\left(x_0\mathrm{cos}(\nu_lt)(1
  -\mathrm{sin}(\nu_lt)),
     y_0\mathrm{cos}(\nu_lt){\mathrm{sin}}(\nu_lt)\right)$,
  where $x_0=y_0=4l_\perp$ and $\nu_l=0.74 (\omega/2\pi)$ is the angular frequency of the
  moving obstacle.  We solve Eq.~(\ref{eqn_gpe}) on a $300^3$ grid~\footnote{We solve Eq.~(\ref{eqn_gpe}) numerically using a 4th
  order Runge Kutta (${\rm{Error}} =O(\Delta t ^4)$) on a discrete grid with spatial step
$\Delta x=\Delta y=\Delta z=0.1 l_\perp$, and time step $\Delta t=0.001 \omega^{-1}$.} with $g=16000 l_\perp\hbar\omega$, 
  where $l_\perp=\sqrt{\hbar/m \omega}$ is the
  harmonic oscillator length, and $\mu=30 \hbar \omega$.  For a trapping frequency of
  $\omega=2\pi\times150$ Hz this corresponds to 
  approximately $21,000$ $^{87}$Rb atoms.   All
  results presented in this paper will be expressed in these units.   
  We choose an obstacle with fixed width ($d=l_\perp/2$) and slowly increase its amplitude $V_0$
  linearly with time ($V_0 = 1.5t$) until it reaches a maximum value at the time where
  it is removed; in this way
  disruptive shock waves are minimized. The total stir time is
  $t_{\mathrm{stir}}=17.1\omega^{-1} $, by which point the beam has undergone almost two
  full oscillations of the figure-eight path (see the left inset of Fig.~\ref{fig:allen_fig3}
  (bottom)) and $V_0$ has reached its maximum value of
  $\approx25.7\hbar\omega$.

Our figure-eight stirring path minimizes the net transfer of angular momentum to the condensate, in comparison to the circular path which imparts angular momentum around the stirring axis and results in an ordered array of vortices arranged in a lattice configuration~\citep{madison_chevy_00,raman_aboshaeer_01,aboshaeer_raman_01,hodby_hechenblaikner_01}. 
   An added advantage to the figure-eight path is that the obstacle
   generates a range of vortex lengths through the whole extent of the condensate (across varying
   axial width), allowing for more bending and tangling of vortex lines than the simpler circular
   stirring.  During the stirring, vortices are nucleated by the obstacle and form a wake behind it.  When the obstacle is removed at time $t =
t_{\rm{stir}}$, what is left is a tangle of reconnecting vortices, as shown in
Fig.~\ref{fig:allen_fig1} (left)~\footnote{See time evolution of isosurfaces in the
  Supplemental
Material provided.}.  The surface of the condensate (spherical when unperturbed) is made uneven by large density waves created by the obstacle and by the motion and reconnections of vortices~\citep{zuccher_caliari_13}. 

 It is difficult to determine by visual inspection how isotropic a vortex tangle is~\citep{white_barenghi_10}.  To measure it precisely, we compute the
  probability density function (normalised histogram, or PDF for short) of the velocity
  components $v_x$, $v_y$ and $v_z$.  The velocity is computed directly from the definition ${\bf{v}}({\bf{r}}) = (\phi^* \nabla \phi - \phi \nabla \phi^*)(2i|\phi|^2).$

  Figure~\ref{fig:allen_fig1} (right) shows such PDFs just after the stirrer has been removed.  The
  good overlap of these velocity PDFs confirms the isotropy of the turbulent velocity
  field.  Moreover, the high degree of symmetry for the PDFs about $v=0$ confirms that negligible linear momentum is imparted in all directions.
For comparison, we also include 
  the Gaussian PDF (gPDF) of each velocity component (correspondingly colored dashed lines), given by
   \begin{eqnarray}
    \label{gpdfs}
   {\rm{gPDF}}(v_i) = \frac{1}{\sigma_i\sqrt{2\pi}}{\rm{exp}}\left(\frac{-(v_i - \tilde \mu_i)^2}{2\sigma_i^2}\right),
    \end{eqnarray}
  where $\sigma_i$ and $\tilde \mu_i$ are the standard deviation and the mean.
  Figure~\ref{fig:allen_fig1} (right) thus also confirms the non-Gaussian (hence non-classical) nature of the velocity PDFs due to the quantised nature of the vortices~\citep{barenghi_lvov_13,paoletti_fisher_08,white_barenghi_10,paoletti_lathrop_11}.  

\begin{figure}[h!]
\centering{
\includegraphics[clip, width=4cm]{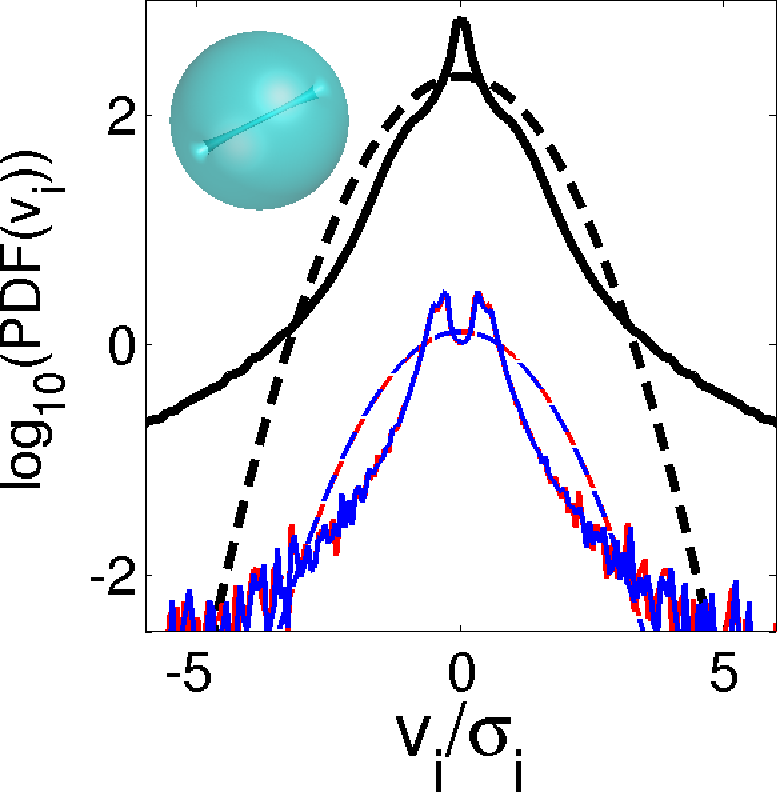}
\hspace{0.3cm}
\includegraphics[clip, width=4.01cm]{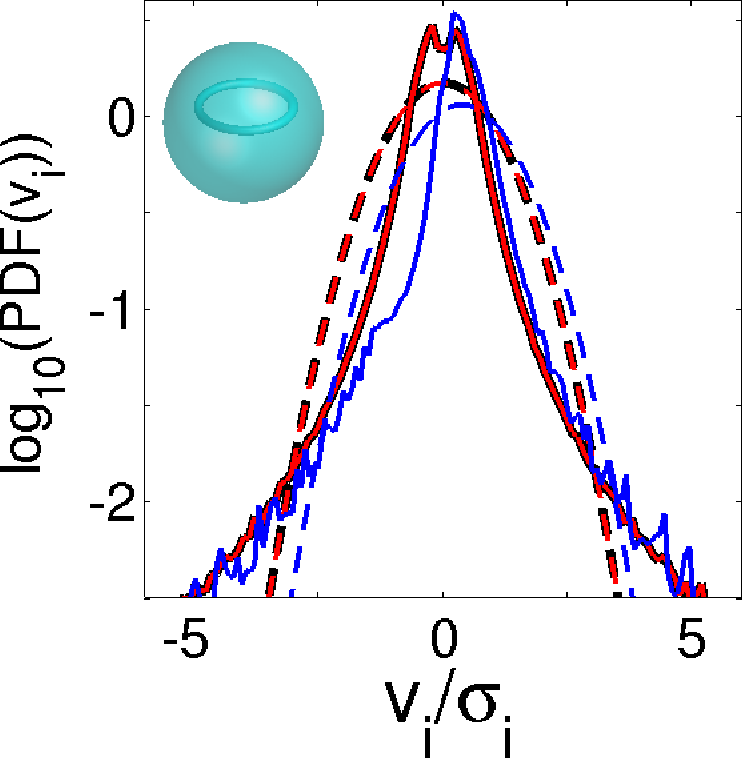}
}
\caption{PDFs of velocity components $v_x$ (black), $v_y$ (red) and $v_z$ (blue) of a single vortex line oriented in the $y-z$ direction (left) and a vortex ring in the $x-y$ plane (right).  Insets show corresponding isosurfaces of the condensate density.}
\label{fig:allen_fig2}
\end{figure}

We have also considered (see Fig.~\ref{fig:allen_fig2}) the velocity PDFs for a straight vortex (left) as well
as a vortex ring (right).  In addition to the non-Gaussian nature, these reveal
non-isotropic velocity PDFs, hence substantiating our claim about the importance of the
isotropy of our generated tangle in Fig.~\ref{fig:allen_fig1} (left).  Although the tangle
shown in Fig.~\ref{fig:allen_fig1}
contains relatively few vortices, the resulting velocity distribution shown in
Fig.~\ref{fig:allen_fig1} (right) resembles the isotropic velocity distributions of a dense vortex tangle in superfluid helium~\citep{baggaley_barenghi_11}.

A further measure to assess isotropy (also routinely used in turbulent superfluid He systems)~\citep{donnelly_91}
is based on projected line length $L_x, L_y$ and $L_z$ in the $x$, $y$ and $z$-directions.  An isotropic vortex configuration will have $L_x/L \approx L_y/L
  \approx L_z/L$ where the total line length $L$.  To determine the vortex length, we find all grid cells where the density has a minimum and
  the phase around that grid point changes by $2\pi$.  We restrict the calculation to within $78\%$ of the Thomas-Fermi radius
  $R_{\rm{TF}} = \sqrt{2\mu}\,l_\perp$ to avoid artefacts arising from the low density edge of the condensate.

\begin{figure}[h!]
    \centering{
      \includegraphics[width=4.1cm]{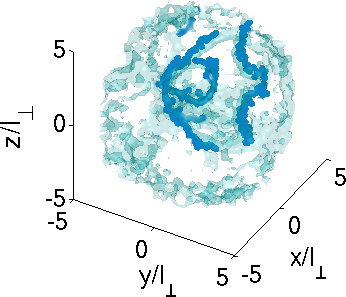} 
      \hspace{0.1cm}
      \includegraphics[width = 4.2cm]{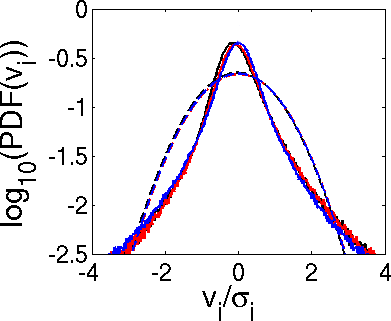} 
    }
    \caption{(Color online) Left: Density isosurface at time $\omega t = 25.7$ (as in Fig.~\ref{fig:allen_fig1}) showing a few remaining
    vortices during the decay of the dense vortex tangle.  Right: Corresponding velocity PDFs
for components $v_x$ (black), $v_y$ (red) and  $v_z$ (blue) (as in Fig.~\ref{fig:allen_fig1}).}
\label{fig:allen_fig4}
  \end{figure}

  Figure~\ref{fig:allen_fig3} (top and intermediate) give condensate density and phase slices 
  (at $z=0$) during the stirring
  ($t<t_{\rm{stir}}$, (A)), just 
  after removing the stirrer ($t\approx t_{\rm{stir}}$, (B)) and at a much later time (C).  Singularities in the phase plot indicate the presence of a vortex; the charge of a vortex can be inferred from the direction of the phase winding with both positive and negatively charged vortices present in these images.  Additionally, these phase plots indicate the presence of nonlinear waves in this system, confirming that this stirring mechanism produces a highly disturbed velocity field.  In C,
  the tangle is less dense than in A and B as most of the vortices have decayed to the
  edge of the condensate.  The bottom part of the figure shows the corresponding projected vortex lengths.  Following a
  near uniform 
  increase of $L$ during stirring ($t<t_{\rm{stir}}$), all vortex lengths $L_x, L_y$ and $L_z$ decay
  together after the stirrer has been removed ($t>t_{\rm{stir}}$), indicating that the vortex
  configuration maintains a high degree of isotropy during its decay.  This is further
  confirmed by inspection of the velocity PDFs at later times (see
  Fig.~\ref{fig:allen_fig4}).  
  
  Figure~\ref{fig:allen_fig3} also shows that during the decay of $L$, all three directional projections oscillate.  
  This is an artifact of the method for calculating the vortex length, which 
  measures it within a fixed spherical volume, $(4\pi/3) R_{\rm{cut}}^3$, where $R_{\rm{cut}} = 0.78R_{\rm{TF}}$.  
  The entire condensate undergoes volume oscillations~\citep{white_barenghi_10},
   and vortices move in and out of the region where the length is determined.  The
   dominant frequency of this linelength oscillation is approximately $2.2\omega$
   which is close to the frequency of
  the monopole mode ($\omega_{osc} = \sqrt{5} \omega$ in the TF approximation~\footnote{It is interesting to
    note that the signature of this mode appears clearly (and only slightly shifted from the
    true monopole mode frequency) from a
  random configuration of vortices, in agreement with R.P. Teles {\emph{et al.}} Phys. Rev. A 88, 053613 (2013) for a single
vortex in the centre of a harmonically trapped condensate.}) of a harmonically-trapped
  BEC~\citep{pethick_smith_book_02,pitaevskii_stringari_book_03}.  It is worth remarking
  that the widths of the condensate in all three directions oscillate in phase
  with each other, in agreement with the
  excitation of this mode~\footnote{We also see evidence of a second frequency appearing which is an exact multiple of the first.}.  Further analysis
  shows that the magnitude of these oscillations increase with stir time.  

  In order to relate vortex linelength to an experimentally observable quantity, such as volume, we
  also measure the norm in the measurement volume $(4\pi/3)R_{\rm{cut}}^3$ and compare this to a condensate of the same total atom number containing no vortices -
  see the right inset of Fig.~\ref{fig:allen_fig3} (bottom).  
  During the stirring, we find that as the linelength increases, the displaced  density also increases,
  with both decreasing when the stirrer is removed.  The displaced density additionally undergoes
  oscillations and these are found to be of the same frequency as those of the vortex length.  
  Since the total atom number remains constant, we infer that the volume of the condensate
  increases to accommodate the vortices.  This effect is visible experimentally~\citep{bagnato_private}
  and can monitored by measuring the atom number within
  a specified radius of the BEC~\citep{jo_choi_97}.  

 In conclusion, we have proposed a novel stirring protocol in ultracold atomic gases, aimed at generating an
   isotropic vortex tangle.  Unlike the usual circular path of stirring (which eventually leads to
    an ordered vortex lattice), and linear sweeping (which
   generates vortex dipoles and can be expected to heavily excite the condensate dipole mode), stirring a spherical 
  condensate along a planar figure-eight path, via a laser-induced obstacle
  aligned along a given axis, results in the generation of an isotropic vortex tangle.  Cessation of the
  stirring leads to the subsequent isotropic decay of the tangle.  As such, this
  stirring protocol represents an efficient route to experimentally generate dense
  isotropic vortex tangles in trapped Bose-Einstein condensates and is a further step towards efficient generation of quantum turbulence in these systems.\\ 
  
  We gratefully acknowledge A. C. White for many useful discussions.
    AJA, NPP and CFB acknowledge funding from the EPSRC (Grant number: EP/I019413/1).     
    
    \appendix
     

\begin{thebibliography}{42}
\expandafter\ifx\csname natexlab\endcsname\relax\def\natexlab#1{#1}\fi
\expandafter\ifx\csname bibnamefont\endcsname\relax
  \def\bibnamefont#1{#1}\fi
\expandafter\ifx\csname bibfnamefont\endcsname\relax
  \def\bibfnamefont#1{#1}\fi
\expandafter\ifx\csname citenamefont\endcsname\relax
  \def\citenamefont#1{#1}\fi
\expandafter\ifx\csname url\endcsname\relax
  \def\url#1{\texttt{#1}}\fi
\expandafter\ifx\csname urlprefix\endcsname\relax\def\urlprefix{URL }\fi
\providecommand{\bibinfo}[2]{#2}
\providecommand{\eprint}[2][]{\url{#2}}

\bibitem[{\citenamefont{Frisch}(1995)}]{frisch_95}
\bibinfo{author}{\bibfnamefont{U.}~\bibnamefont{Frisch}},
  \emph{\bibinfo{title}{Turbulence. The legacy of A.N. Kolmogorov.}}
  (\bibinfo{publisher}{Cambridge University Press}, \bibinfo{year}{1995}).

\bibitem[{\citenamefont{Barenghi et~al.}(2001)\citenamefont{Barenghi, Donnelly,
  Vinen, and eds.}}]{barenghi_donnelly_01}
\bibinfo{author}{\bibfnamefont{C.}~\bibnamefont{Barenghi}},
  \bibinfo{author}{\bibfnamefont{R.}~\bibnamefont{Donnelly}},
  \bibinfo{author}{\bibfnamefont{W.}~\bibnamefont{Vinen}}, \bibnamefont{and}
  \bibinfo{author}{\bibnamefont{eds.}}, \emph{\bibinfo{title}{Quantized Vortex
  Dynamics and Superfluid Turbulence}} (\bibinfo{publisher}{Springer, Berlin},
  \bibinfo{year}{2001}).

\bibitem[{\citenamefont{Maurer and Tabeling}(1998)}]{maurer_tabeling_98}
\bibinfo{author}{\bibfnamefont{J.}~\bibnamefont{Maurer}} \bibnamefont{and}
  \bibinfo{author}{\bibfnamefont{P.}~\bibnamefont{Tabeling}},
  \bibinfo{journal}{EPL (Europhysics Letters)} \textbf{\bibinfo{volume}{43}},
  \bibinfo{pages}{29} (\bibinfo{year}{1998}).

\bibitem[{\citenamefont{Salort et~al.}(2010)\citenamefont{Salort, Baudet,
  Castaing, Chabaud, Daviaud, Didelot, Diribarne, Dubrulle, Gagne, Gauthier
  et~al.}}]{salort_baudet_10}
\bibinfo{author}{\bibfnamefont{J.}~\bibnamefont{Salort}},
  \bibinfo{author}{\bibfnamefont{C.}~\bibnamefont{Baudet}},
  \bibinfo{author}{\bibfnamefont{B.}~\bibnamefont{Castaing}},
  \bibinfo{author}{\bibfnamefont{B.}~\bibnamefont{Chabaud}},
  \bibinfo{author}{\bibfnamefont{F.}~\bibnamefont{Daviaud}},
  \bibinfo{author}{\bibfnamefont{T.}~\bibnamefont{Didelot}},
  \bibinfo{author}{\bibfnamefont{P.}~\bibnamefont{Diribarne}},
  \bibinfo{author}{\bibfnamefont{B.}~\bibnamefont{Dubrulle}},
  \bibinfo{author}{\bibfnamefont{Y.}~\bibnamefont{Gagne}},
  \bibinfo{author}{\bibfnamefont{F.}~\bibnamefont{Gauthier}},
  \bibnamefont{et~al.}, \bibinfo{journal}{Physics of Fluids}
  \textbf{\bibinfo{volume}{22}}, \bibinfo{eid}{125102}
   (\bibinfo{year}{2010}).

\bibitem[{\citenamefont{Barenghi et~al.}(2013)\citenamefont{Barenghi, L'vov,
  and Roche}}]{barenghi_lvov_13}
\bibinfo{author}{\bibfnamefont{C.}~\bibnamefont{Barenghi}},
  \bibinfo{author}{\bibfnamefont{V.}~\bibnamefont{L'vov}}, \bibnamefont{and}
  \bibinfo{author}{\bibfnamefont{P.-E.} \bibnamefont{Roche}}
  (\bibinfo{year}{2013}), \bibinfo{note}{arXiv:1306.6248 [cond-mat.other]}.

\bibitem{allen_parker_13}
  \bibinfo{author}{\bibfnamefont{A.~J.}~\bibnamefont{Allen}}, 
\bibinfo{author}{\bibfnamefont{N.~G.}~\bibnamefont{Parker}}, 
\bibinfo{author}{\bibfnamefont{N.~P.}~\bibnamefont{Proukakis}} \bibnamefont{and}
  \bibinfo{author}{\bibfnamefont{C.~F.}~\bibnamefont{Barenghi}},
  (\bibinfo{year}{2013}), \bibinfo{note}{arXiv:1302.7176 [cond-mat.quant-gas]}.

\bibitem[{\citenamefont{Anderson et~al.}(1995)\citenamefont{Anderson, Ensher,
  Matthews, and Wiemann}}]{anderson_ensher_95}
\bibinfo{author}{\bibfnamefont{M.~H.} \bibnamefont{Anderson}},
  \bibinfo{author}{\bibfnamefont{J.}~\bibnamefont{Ensher}},
  \bibinfo{author}{\bibfnamefont{M.~R.} \bibnamefont{Matthews}},
  \bibnamefont{and} \bibinfo{author}{\bibfnamefont{C.~E.}
  \bibnamefont{Wiemann}}, \bibinfo{journal}{Science}
  \textbf{\bibinfo{volume}{269}}, \bibinfo{pages}{198} (\bibinfo{year}{1995}).

\bibitem[{\citenamefont{Davis et~al.}(1995)\citenamefont{Davis, Mewes, Andrews,
  van Druten, Durfee, Kurn, and Ketterle}}]{davis_mewes_95}
\bibinfo{author}{\bibfnamefont{K.~B.} \bibnamefont{Davis}},
  \bibinfo{author}{\bibfnamefont{M.~O.} \bibnamefont{Mewes}},
  \bibinfo{author}{\bibfnamefont{M.~R.} \bibnamefont{Andrews}},
  \bibinfo{author}{\bibfnamefont{N.~J.} \bibnamefont{van Druten}},
  \bibinfo{author}{\bibfnamefont{D.~S.} \bibnamefont{Durfee}},
  \bibinfo{author}{\bibfnamefont{D.~M.} \bibnamefont{Kurn}}, \bibnamefont{and}
  \bibinfo{author}{\bibfnamefont{W.}~\bibnamefont{Ketterle}},
  \bibinfo{journal}{Phys. Rev. Lett.} \textbf{\bibinfo{volume}{75}},
  \bibinfo{pages}{3969} (\bibinfo{year}{1995}).

\bibitem[{\citenamefont{Bradley et~al.}(1997)\citenamefont{Bradley, Sackett,
  Tollett, and Hulet}}]{bradley_sackett_97}
\bibinfo{author}{\bibfnamefont{C.~C.} \bibnamefont{Bradley}},
  \bibinfo{author}{\bibfnamefont{C.~A.} \bibnamefont{Sackett}},
  \bibinfo{author}{\bibfnamefont{J.~J.} \bibnamefont{Tollett}},
  \bibnamefont{and} \bibinfo{author}{\bibfnamefont{R.~G.} \bibnamefont{Hulet}},
  \bibinfo{journal}{Phys. Rev. Lett.} \textbf{\bibinfo{volume}{79}},
  \bibinfo{pages}{1170} (\bibinfo{year}{1997}).

\bibitem[{\citenamefont{Madison et~al.}(2000)\citenamefont{Madison, Chevy,
  Wohlleben, and Dalibard}}]{madison_chevy_00}
\bibinfo{author}{\bibfnamefont{K.~W.} \bibnamefont{Madison}},
  \bibinfo{author}{\bibfnamefont{F.}~\bibnamefont{Chevy}},
  \bibinfo{author}{\bibfnamefont{W.}~\bibnamefont{Wohlleben}},
  \bibnamefont{and} \bibinfo{author}{\bibfnamefont{J.}~\bibnamefont{Dalibard}},
  \bibinfo{journal}{Phys. Rev. Lett.} \textbf{\bibinfo{volume}{84}},
  \bibinfo{pages}{806} (\bibinfo{year}{2000}).



\bibitem[{\citenamefont{Hodby et~al.}(2001)\citenamefont{Hodby, Hechenblaikner,
  Hopkins, Marag\`o, and Foot}}]{hodby_hechenblaikner_01}
\bibinfo{author}{\bibfnamefont{E.}~\bibnamefont{Hodby}},
  \bibinfo{author}{\bibfnamefont{G.}~\bibnamefont{Hechenblaikner}},
  \bibinfo{author}{\bibfnamefont{S.~A.} \bibnamefont{Hopkins}},
  \bibinfo{author}{\bibfnamefont{O.~M.} \bibnamefont{Marag\`o}},
  \bibnamefont{and} \bibinfo{author}{\bibfnamefont{C.~J.} \bibnamefont{Foot}},
  \bibinfo{journal}{Phys. Rev. Lett.} \textbf{\bibinfo{volume}{88}},
  \bibinfo{pages}{010405} (\bibinfo{year}{2001}).

\bibitem[{\citenamefont{Abo-Shaeer et~al.}(2002)\citenamefont{Abo-Shaeer,
  Raman, and Ketterle}}]{abo-shaeer_raman_02}
\bibinfo{author}{\bibfnamefont{J.~R.} \bibnamefont{Abo-Shaeer}},
  \bibinfo{author}{\bibfnamefont{C.}~\bibnamefont{Raman}}, \bibnamefont{and}
  \bibinfo{author}{\bibfnamefont{W.}~\bibnamefont{Ketterle}},
  \bibinfo{journal}{Phys. Rev. Lett.} \textbf{\bibinfo{volume}{88}},
  \bibinfo{pages}{070409} (\bibinfo{year}{2002}).

\bibitem[{\citenamefont{Abo-Shaeer et~al.}(2001)\citenamefont{Abo-Shaeer,
  Raman, Vogels, and Ketterle}}]{aboshaeer_raman_01}
\bibinfo{author}{\bibfnamefont{J.~R.} \bibnamefont{Abo-Shaeer}},
  \bibinfo{author}{\bibfnamefont{C.}~\bibnamefont{Raman}},
  \bibinfo{author}{\bibfnamefont{J.~M.} \bibnamefont{Vogels}},
  \bibnamefont{and} \bibinfo{author}{\bibfnamefont{W.}~\bibnamefont{Ketterle}},
  \bibinfo{journal}{Science} \textbf{\bibinfo{volume}{292}},
  \bibinfo{pages}{476} (\bibinfo{year}{2001}).

\bibitem[{\citenamefont{Madison et~al.}(2001)\citenamefont{Madison, Chevy,
  Bretin, and Dalibard}}]{madison_chevy_01}
\bibinfo{author}{\bibfnamefont{K.~W.} \bibnamefont{Madison}},
  \bibinfo{author}{\bibfnamefont{F.}~\bibnamefont{Chevy}},
  \bibinfo{author}{\bibfnamefont{V.}~\bibnamefont{Bretin}}, \bibnamefont{and}
  \bibinfo{author}{\bibfnamefont{J.}~\bibnamefont{Dalibard}},
  \bibinfo{journal}{Phys. Rev. Lett.} \textbf{\bibinfo{volume}{86}},
  \bibinfo{pages}{4443} (\bibinfo{year}{2001}).%

\bibitem[{\citenamefont{Raman et~al.}(2001{\natexlab{a}})\citenamefont{Raman,
  Abo-Shaeer, Vogels, Xu, and Ketterle}}]{raman_aboshaeer_01}
\bibinfo{author}{\bibfnamefont{C.}~\bibnamefont{Raman}},
  \bibinfo{author}{\bibfnamefont{J.~R.} \bibnamefont{Abo-Shaeer}},
  \bibinfo{author}{\bibfnamefont{J.~M.} \bibnamefont{Vogels}},
  \bibinfo{author}{\bibfnamefont{K.}~\bibnamefont{Xu}}, \bibnamefont{and}
  \bibinfo{author}{\bibfnamefont{W.}~\bibnamefont{Ketterle}},
  \bibinfo{journal}{Phys. Rev. Lett.} \textbf{\bibinfo{volume}{87}},
  \bibinfo{pages}{210402} (\bibinfo{year}{2001}{\natexlab{a}}).

\bibitem[{\citenamefont{Neely et~al.}(2010)\citenamefont{Neely, Samson,
  Bradley, Davis, and Anderson}}]{neely_samson_10}
\bibinfo{author}{\bibfnamefont{T.~W.} \bibnamefont{Neely}},
  \bibinfo{author}{\bibfnamefont{E.~C.} \bibnamefont{Samson}},
  \bibinfo{author}{\bibfnamefont{A.~S.} \bibnamefont{Bradley}},
  \bibinfo{author}{\bibfnamefont{M.~J.} \bibnamefont{Davis}}, \bibnamefont{and}
  \bibinfo{author}{\bibfnamefont{B.~P.} \bibnamefont{Anderson}},
  \bibinfo{journal}{Phys. Rev. Lett.} \textbf{\bibinfo{volume}{104}} \bibinfo{pages}{160401}
  (\bibinfo{year}{2010}).

\bibitem[{\citenamefont{Shin et~al.}(2004)\citenamefont{Shin, Saba,
  Vengalattore, Pasquini, Sanner, Leanhardt, Prentiss, Pritchard, and
  Ketterle}}]{shin_saba_04}
\bibinfo{author}{\bibfnamefont{Y.}~\bibnamefont{Shin}},
  \bibinfo{author}{\bibfnamefont{M.}~\bibnamefont{Saba}},
  \bibinfo{author}{\bibfnamefont{M.}~\bibnamefont{Vengalattore}},
  \bibinfo{author}{\bibfnamefont{T.~A.} \bibnamefont{Pasquini}},
  \bibinfo{author}{\bibfnamefont{C.}~\bibnamefont{Sanner}},
  \bibinfo{author}{\bibfnamefont{A.~E.} \bibnamefont{Leanhardt}},
  \bibinfo{author}{\bibfnamefont{M.}~\bibnamefont{Prentiss}},
  \bibinfo{author}{\bibfnamefont{D.~E.} \bibnamefont{Pritchard}},
  \bibnamefont{and} \bibinfo{author}{\bibfnamefont{W.}~\bibnamefont{Ketterle}},
  \bibinfo{journal}{Phys. Rev. Lett.} \textbf{\bibinfo{volume}{93}},
  \bibinfo{pages}{160406} (\bibinfo{year}{2004}).

\bibitem[{\citenamefont{Matthews et~al.}(1999)\citenamefont{Matthews, Anderson,
  Haljan, Hall, Wieman, and Cornell}}]{matthews_anderson_99}
\bibinfo{author}{\bibfnamefont{M.~R.} \bibnamefont{Matthews}},
  \bibinfo{author}{\bibfnamefont{B.~P.} \bibnamefont{Anderson}},
  \bibinfo{author}{\bibfnamefont{P.~C.} \bibnamefont{Haljan}},
  \bibinfo{author}{\bibfnamefont{D.~S.} \bibnamefont{Hall}},
  \bibinfo{author}{\bibfnamefont{C.~E.} \bibnamefont{Wieman}},
  \bibnamefont{and} \bibinfo{author}{\bibfnamefont{E.~A.}
  \bibnamefont{Cornell}}, \bibinfo{journal}{Phys. Rev. Lett.}
  \textbf{\bibinfo{volume}{83}}, \bibinfo{pages}{2498} (\bibinfo{year}{1999}).

\bibitem[{\citenamefont{Anderson et~al.}(2000)\citenamefont{Anderson, Haljan,
  Wieman, and Cornell}}]{anderson_haljan_00}
\bibinfo{author}{\bibfnamefont{B.~P.} \bibnamefont{Anderson}},
  \bibinfo{author}{\bibfnamefont{P.~C.} \bibnamefont{Haljan}},
  \bibinfo{author}{\bibfnamefont{C.~E.} \bibnamefont{Wieman}},
  \bibnamefont{and} \bibinfo{author}{\bibfnamefont{E.~A.}
  \bibnamefont{Cornell}}, \bibinfo{journal}{Phys. Rev. Lett.}
  \textbf{\bibinfo{volume}{85}}, \bibinfo{pages}{2857} (\bibinfo{year}{2000}).

\bibitem[{\citenamefont{Schweikhard et~al.}(2004)\citenamefont{Schweikhard,
  Coddington, Engels, Tung, and Cornell}}]{schweikhard_coddington_04}
\bibinfo{author}{\bibfnamefont{V.}~\bibnamefont{Schweikhard}},
  \bibinfo{author}{\bibfnamefont{I.}~\bibnamefont{Coddington}},
  \bibinfo{author}{\bibfnamefont{P.}~\bibnamefont{Engels}},
  \bibinfo{author}{\bibfnamefont{S.}~\bibnamefont{Tung}}, \bibnamefont{and}
  \bibinfo{author}{\bibfnamefont{E.~A.} \bibnamefont{Cornell}},
  \bibinfo{journal}{Phys. Rev. Lett.} \textbf{\bibinfo{volume}{93}},
  \bibinfo{pages}{210403} (\bibinfo{year}{2004}).

\bibitem[{\citenamefont{Freilich et~al.}(2010)\citenamefont{Freilich, Bianchi,
  Kaufman, Langin, and Hall}}]{freilich_bianchi_10}
\bibinfo{author}{\bibfnamefont{D.~V.} \bibnamefont{Freilich}},
  \bibinfo{author}{\bibfnamefont{D.~M.} \bibnamefont{Bianchi}},
  \bibinfo{author}{\bibfnamefont{A.~M.} \bibnamefont{Kaufman}},
  \bibinfo{author}{\bibfnamefont{T.~K.} \bibnamefont{Langin}},
  \bibnamefont{and} \bibinfo{author}{\bibfnamefont{D.~S.} \bibnamefont{Hall}},
  \bibinfo{journal}{Science} \textbf{\bibinfo{volume}{329}},
  \bibinfo{pages}{1182} (\bibinfo{year}{2010}).

\bibitem[{\citenamefont{Henn et~al.}(2009)\citenamefont{Henn, Seman, Roati,
  Magalh\~aes, and Bagnato}}]{henn_seman_09}
\bibinfo{author}{\bibfnamefont{E.~A.~L.} \bibnamefont{Henn}},
  \bibinfo{author}{\bibfnamefont{J.~A.} \bibnamefont{Seman}},
  \bibinfo{author}{\bibfnamefont{G.}~\bibnamefont{Roati}},
  \bibinfo{author}{\bibfnamefont{K.~M.~F.} \bibnamefont{Magalh\~aes}},
  \bibnamefont{and} \bibinfo{author}{\bibfnamefont{V.~S.}
  \bibnamefont{Bagnato}}, \bibinfo{journal}{Phys. Rev. Lett.}
  \textbf{\bibinfo{volume}{103}}, \bibinfo{pages}{045301}
  (\bibinfo{year}{2009}).

\bibitem[{\citenamefont{Seman et~al.}(2011)\citenamefont{Seman, Henn, Shiozaki,
  Roati, Poveda-Cuevas, Magalhães, Yukalov, Tsubota, Kobayashi, Kasamatsu
  et~al.}}]{seman_henn_11}
\bibinfo{author}{\bibfnamefont{J.}~\bibnamefont{Seman}},
  \bibinfo{author}{\bibfnamefont{E.}~\bibnamefont{Henn}},
  \bibinfo{author}{\bibfnamefont{R.}~\bibnamefont{Shiozaki}},
  \bibinfo{author}{\bibfnamefont{G.}~\bibnamefont{Roati}},
  \bibinfo{author}{\bibfnamefont{F.}~\bibnamefont{Poveda-Cuevas}},
  \bibinfo{author}{\bibfnamefont{K.}~\bibnamefont{Magalhães}},
  \bibinfo{author}{\bibfnamefont{V.}~\bibnamefont{Yukalov}},
  \bibinfo{author}{\bibfnamefont{M.}~\bibnamefont{Tsubota}},
  \bibinfo{author}{\bibfnamefont{M.}~\bibnamefont{Kobayashi}},
  \bibinfo{author}{\bibfnamefont{K.}~\bibnamefont{Kasamatsu}},
  \bibnamefont{et~al.}, \bibinfo{journal}{Laser Physics Letters}
  \textbf{\bibinfo{volume}{8}}, \bibinfo{pages}{691} (\bibinfo{year}{2011}).

\bibitem[{\citenamefont{Shiozaki et~al.}(2011)\citenamefont{Shiozaki, Telles,
  Yukalov, and Bagnato}}]{shiozaki_telles_11}
\bibinfo{author}{\bibfnamefont{R.}~\bibnamefont{Shiozaki}},
  \bibinfo{author}{\bibfnamefont{G.}~\bibnamefont{Telles}},
  \bibinfo{author}{\bibfnamefont{V.}~\bibnamefont{Yukalov}}, \bibnamefont{and}
  \bibinfo{author}{\bibfnamefont{V.}~\bibnamefont{Bagnato}},
  \bibinfo{journal}{Laser Physics Letters} \textbf{\bibinfo{volume}{8}},
  \bibinfo{pages}{393} (\bibinfo{year}{2011}).

\bibitem[{\citenamefont{Nore et~al.}(1997)\citenamefont{Nore, Abid, and
  Brachet}}]{nore_abid_07}
\bibinfo{author}{\bibfnamefont{C.}~\bibnamefont{Nore}},
  \bibinfo{author}{\bibfnamefont{M.}~\bibnamefont{Abid}}, \bibnamefont{and}
  \bibinfo{author}{\bibfnamefont{M.~E.} \bibnamefont{Brachet}},
  \bibinfo{journal}{Physics of Fluids} \textbf{\bibinfo{volume}{9}},
  \bibinfo{pages}{2644} (\bibinfo{year}{1997}).

\bibitem[{\citenamefont{Berloff and Svistunov}(2002)}]{berloff_svistunov_02}
\bibinfo{author}{\bibfnamefont{N.~G.} \bibnamefont{Berloff}} \bibnamefont{and}
  \bibinfo{author}{\bibfnamefont{B.~V.} \bibnamefont{Svistunov}},
  \bibinfo{journal}{Phys. Rev. A} \textbf{\bibinfo{volume}{66}},
  \bibinfo{pages}{013603} (\bibinfo{year}{2002}).

\bibitem[{\citenamefont{Yepez et~al.}(2009)\citenamefont{Yepez, Vahala, Vahala,
  and Soe}}]{yepez_vahala_09}
\bibinfo{author}{\bibfnamefont{J.}~\bibnamefont{Yepez}},
  \bibinfo{author}{\bibfnamefont{G.}~\bibnamefont{Vahala}},
  \bibinfo{author}{\bibfnamefont{L.}~\bibnamefont{Vahala}}, \bibnamefont{and}
  \bibinfo{author}{\bibfnamefont{M.}~\bibnamefont{Soe}},
  \bibinfo{journal}{Phys. Rev. Lett.} \textbf{\bibinfo{volume}{103}},
  \bibinfo{pages}{084501} (\bibinfo{year}{2009}).

\bibitem[{\citenamefont{White et~al.}(2010)\citenamefont{White, Barenghi,
  Proukakis, Youd, and Wacks}}]{white_barenghi_10}
\bibinfo{author}{\bibfnamefont{A.~C.} \bibnamefont{White}},
  \bibinfo{author}{\bibfnamefont{C.~F.} \bibnamefont{Barenghi}},
  \bibinfo{author}{\bibfnamefont{N.~P.} \bibnamefont{Proukakis}},
  \bibinfo{author}{\bibfnamefont{A.~J.} \bibnamefont{Youd}}, \bibnamefont{and}
  \bibinfo{author}{\bibfnamefont{D.~H.} \bibnamefont{Wacks}},
  \bibinfo{journal}{Phys. Rev. Lett.} \textbf{\bibinfo{volume}{104}},
  \bibinfo{pages}{075301} (\bibinfo{year}{2010}).

\bibitem[{\citenamefont{Kobayashi and Tsubota}(2007)}]{kobayashi_tsubota_07}
\bibinfo{author}{\bibfnamefont{M.}~\bibnamefont{Kobayashi}} \bibnamefont{and}
  \bibinfo{author}{\bibfnamefont{M.}~\bibnamefont{Tsubota}},
  \bibinfo{journal}{Phys. Rev. A} \textbf{\bibinfo{volume}{76}},
  \bibinfo{pages}{045603} (\bibinfo{year}{2007}).

\bibitem{gaunt_schmidutz_13}
\bibinfo{author}{\bibfnamefont{A.~L.}~\bibnamefont{Gaunt}},
\bibinfo{author}{\bibfnamefont{T.~F.}~\bibnamefont{Schmidutz}}, 
\bibinfo{author}{\bibfnamefont{I.}~\bibnamefont{Gotlibovych}},
\bibinfo{author}{\bibfnamefont{R.~P}~\bibnamefont{Smith}}
\bibnamefont{and}
\bibinfo{author}{\bibfnamefont{Z.}~\bibnamefont{Hadzibabic}},
\bibinfo{journal}{Phys. Rev. Lett.} \textbf{\bibinfo{volume}{110}}, 
\bibinfo{pages}{200406} (\bibinfo{year}{2013}).



\bibitem[{\citenamefont{Frisch et~al.}(1992)\citenamefont{Frisch, Pomeau, and
  Rica}}]{frisch_pomeau_92}
\bibinfo{author}{\bibfnamefont{T.}~\bibnamefont{Frisch}},
  \bibinfo{author}{\bibfnamefont{Y.}~\bibnamefont{Pomeau}}, \bibnamefont{and}
  \bibinfo{author}{\bibfnamefont{S.}~\bibnamefont{Rica}},
  \bibinfo{journal}{Phys. Rev. Lett.} \textbf{\bibinfo{volume}{69}},
  \bibinfo{pages}{1644} (\bibinfo{year}{1992}).

\bibitem{caradoc-davis_ballagh_99}
\bibinfo{author}{\bibnamefont{B.~M.}~\bibnamefont{Caradoc-Davies}},
\bibinfo{author}{\bibnamefont{R.~J.}~\bibnamefont{Ballagh}},
\bibnamefont{and}
\bibinfo{author}{\bibnamefont{K.}~\bibnamefont{Burnett}},
\bibinfo{journal}{Phys. Rev. Lett.} \textbf{\bibinfo{volume}{83}},
\bibinfo{page}{895} (\bibinfo{year}{1999}).

\bibitem[{\citenamefont{Raman et~al.}(2001{\natexlab{b}})\citenamefont{Raman,
  K\"{o}hl, Onofrio, Durfee, Kuklewicz, Hadzibabic, and Ketterle
 and Ketterle}}]{raman_kohl_99}
\bibinfo{author}{\bibfnamefont{C.}~\bibnamefont{Raman}},
  \bibinfo{author}{\bibfnamefont{M.} \bibnamefont{K\"{o}hl}},
  \bibinfo{author}{\bibfnamefont{R.} \bibnamefont{Onofrio}},
  \bibinfo{author}{\bibfnamefont{D.~S.}~\bibnamefont{Durfee}}, 
 \bibinfo{author}{\bibfnamefont{C.~E.}~\bibnamefont{Kuklewicz}}, 
 \bibinfo{author}{\bibfnamefont{Z.}~\bibnamefont{Hadzibabic}}, 
\bibnamefont{and}
  \bibinfo{author}{\bibfnamefont{W.}~\bibnamefont{Ketterle}},
  \bibinfo{journal}{Phys. Rev. Lett.} \textbf{\bibinfo{volume}{83}},
  \bibinfo{pages}{2502} (\bibinfo{year}{1999}{\natexlab{b}}).

\bibitem{kevrekidis_frantzeskakis_08}
\bibinfo{author}{\bibfnamefont{G.~P.}~\bibnamefont{Kevrekidis}},
  \bibinfo{author}{\bibfnamefont{D.~J.}~\bibnamefont{Frantzeskakis}},
  \bibinfo{author}{\bibfnamefont{R.}~\bibnamefont{Carretero-González}}
  \bibinfo{author}{\bibnamefont{(Eds.)}}, \emph{\bibinfo{title}{Emergent Nonlinear Phenomena in Bose-Einstein Condensates, Theory and Experiment }} (\bibinfo{publisher}{Springer Series on Atomic, Optical, and Plasma Physics, Vol. 45},
  \bibinfo{year}{2008}).


\bibitem[{\citenamefont{White et~al.}(2012)\citenamefont{White, Barenghi, and
  Proukakis}}]{white_barenghi_12}
\bibinfo{author}{\bibfnamefont{A.~C.} \bibnamefont{White}},
  \bibinfo{author}{\bibfnamefont{C.~F.} \bibnamefont{Barenghi}},
  \bibnamefont{and} \bibinfo{author}{\bibfnamefont{N.~P.}
  \bibnamefont{Proukakis}}, \bibinfo{journal}{Phys. Rev. A}
  \textbf{\bibinfo{volume}{86}}, \bibinfo{pages}{013635}
  (\bibinfo{year}{2012}).

\bibitem[{\citenamefont{Reeves et~al.}(2012)\citenamefont{Reeves, Anderson, and
  Bradley}}]{reeves_anderson_12}
\bibinfo{author}{\bibfnamefont{M.~T.} \bibnamefont{Reeves}},
  \bibinfo{author}{\bibfnamefont{B.~P.} \bibnamefont{Anderson}},
  \bibnamefont{and} \bibinfo{author}{\bibfnamefont{A.~S.}
  \bibnamefont{Bradley}}, \bibinfo{journal}{Phys. Rev. A}
  \textbf{\bibinfo{volume}{86}}, \bibinfo{pages}{053621}
  (\bibinfo{year}{2012}).

\bibitem[{\citenamefont{Reeves et~al.}(2013)\citenamefont{Reeves, Billam,
  Anderson, and Bradley}}]{reeves_billam_13}
\bibinfo{author}{\bibfnamefont{M.~T.} \bibnamefont{Reeves}},
  \bibinfo{author}{\bibfnamefont{T.~P.} \bibnamefont{Billam}},
  \bibinfo{author}{\bibfnamefont{B.~P.} \bibnamefont{Anderson}},
  \bibnamefont{and} \bibinfo{author}{\bibfnamefont{A.~S.}
  \bibnamefont{Bradley}}, \bibinfo{journal}{Phys. Rev. Lett.}
  \textbf{\bibinfo{volume}{110}}, \bibinfo{pages}{104501}
  (\bibinfo{year}{2013}).

\bibitem[{\citenamefont{Donnelly}(1991)}]{donnelly_91}
\bibinfo{author}{\bibfnamefont{R.~J.} \bibnamefont{Donnelly}},
  \emph{\bibinfo{title}{Quantized Vortices in Helium II}}
  (\bibinfo{publisher}{Cambridge Univ. Press, Cambridge},
  \bibinfo{year}{1991}).

\bibitem[{\citenamefont{Pethick and Smith}(2002)}]{pethick_smith_book_02}
\bibinfo{author}{\bibfnamefont{C.~J.}~\bibnamefont{Pethick}} \bibnamefont{and}
  \bibinfo{author}{\bibfnamefont{H.}~\bibnamefont{Smith}},
  \emph{\bibinfo{title}{Bose-Einstein condensation in dilute gases}}
  (\bibinfo{publisher}{Cambridge University Press}, \bibinfo{year}{2002}).

\bibitem[{\citenamefont{Paoletti et~al.}(2008)\citenamefont{Paoletti, Fisher,
  Sreenivasan, and Lathrop}}]{paoletti_fisher_08}
\bibinfo{author}{\bibfnamefont{M.~S.} \bibnamefont{Paoletti}},
  \bibinfo{author}{\bibfnamefont{M.~E.} \bibnamefont{Fisher}},
  \bibinfo{author}{\bibfnamefont{K.~R.} \bibnamefont{Sreenivasan}},
  \bibnamefont{and} \bibinfo{author}{\bibfnamefont{D.~P.}
  \bibnamefont{Lathrop}}, \bibinfo{journal}{Phys. Rev. Lett.}
  \textbf{\bibinfo{volume}{101}}, \bibinfo{pages}{154501}
  (\bibinfo{year}{2008}).



\bibitem[{\citenamefont{Pitaevskii and
  Stringari}(2003)}]{pitaevskii_stringari_book_03}
\bibinfo{author}{\bibfnamefont{L.~P.} \bibnamefont{Pitaevskii}}
  \bibnamefont{and}
  \bibinfo{author}{\bibfnamefont{S.}~\bibnamefont{Stringari}},
  \emph{\bibinfo{title}{Bose-Einstein Condensation}}
  (\bibinfo{publisher}{Oxford University Press}, \bibinfo{address}{Great
  Clarendon Street, Oxford}, \bibinfo{year}{2003}).

\bibitem{adams_riis_97}
\bibinfo{author}{\bibnamefont{C.~S.}~\bibnamefont{Adams}} \bibnamefont{and}
\bibinfo{author}{\bibnamefont{E.}~\bibnamefont{Riis}},
\bibinfo{journal}{Progress in Quantum Electronics} \textbf{\bibinfo{volume}{21}},
\bibinfo{pages}{1} (\bibinfo{year}{1997}).







\bibitem[{\citenamefont{Zuccher et~al.}(2012)\citenamefont{Zuccher, Caliari,
  Baggaley, and Barenghi}}]{zuccher_caliari_13}
\bibinfo{author}{\bibfnamefont{S.}~\bibnamefont{Zuccher}},
  \bibinfo{author}{\bibfnamefont{M.}~\bibnamefont{Caliari}},
  \bibinfo{author}{\bibfnamefont{A.~W.} \bibnamefont{Baggaley}},
  \bibnamefont{and} \bibinfo{author}{\bibfnamefont{C.~F.}
  \bibnamefont{Barenghi}}, \bibinfo{journal}{Physics of Fluids}
  \textbf{\bibinfo{volume}{24}}, \bibinfo{eid}{125108} (\bibinfo{year}{2012}).

\bibitem{paoletti_lathrop_11}
\bibinfo{author}{\bibfnamefont{M.~S.} \bibnamefont{Paoletti}},
  \bibnamefont{and} \bibinfo{author}{\bibfnamefont{D.~P.}
  \bibnamefont{Lathrop}}, \bibinfo{journal}{Ann. Rev. Cond. Matt. Phys.}
  \textbf{\bibinfo{volume}{2}}, \bibinfo{pages}{213}
  (\bibinfo{year}{2011}).


\bibitem{baggaley_barenghi_11}
\bibinfo{author}{\bibnamefont{A.~W.} \bibnamefont{Baggaley}}, 
\bibnamefont{and}
\bibinfo{author}{\bibnamefont{C.~F.} \bibnamefont{Barenghi}},
\bibinfo{journal}{Phys. Rev. E} \textbf{\bibinfo{volume}{84}},
\bibinfo{pages}{067301}
(\bibinfo{year}{2011}).



%



\bibitem[{\citenamefont{Bagnato}()}]{bagnato_private}
\bibinfo{author}{\bibfnamefont{V.~S.} \bibnamefont{Bagnato}},
  \emph{\bibinfo{title}{Private communication}}.

\bibitem[{\citenamefont{Jo et~al.}(2007)\citenamefont{Jo, Choi, Christensen,
  Pasquini, Lee, Ketterle, and Pritchard}}]{jo_choi_97}
\bibinfo{author}{\bibfnamefont{G.-B.} \bibnamefont{Jo}},
  \bibinfo{author}{\bibfnamefont{J.-H.} \bibnamefont{Choi}},
  \bibinfo{author}{\bibfnamefont{C.~A.} \bibnamefont{Christensen}},
  \bibinfo{author}{\bibfnamefont{T.~A.} \bibnamefont{Pasquini}},
  \bibinfo{author}{\bibfnamefont{Y.-R.} \bibnamefont{Lee}},
  \bibinfo{author}{\bibfnamefont{W.}~\bibnamefont{Ketterle}}, \bibnamefont{and}
  \bibinfo{author}{\bibfnamefont{D.~E.} \bibnamefont{Pritchard}},
  \bibinfo{journal}{Phys. Rev. Lett.} \textbf{\bibinfo{volume}{98}},
  \bibinfo{pages}{180401} (\bibinfo{year}{2007}).

 
\end{thebibliography}
    
\end{document}